\documentclass[aps,prd,twocolumn,showkeys,showpacs,superscriptaddress,groupedaddress,amsfonts]{revtex4-1}
\usepackage{graphicx}  
\usepackage{dcolumn}   
\usepackage{bm}        
\usepackage{amssymb}   
\usepackage{mathrsfs}
\usepackage{amsmath}
\usepackage{txfonts}
\usepackage{multirow}
\begin{document}
	
\title{Dynamical system analysis of Einstein-Skyrme model in a Kantowski-Sachs spacetime }

\author{Sudip Mishra \thanks{corresponding author}}
\email[]{sudipcmiiitmath@gmail.com}
\author{Subenoy Chakraborty}
	
\affiliation{Department of Mathematics, Jadavpur University, Kolkata- 700032, WB, India.}
\date{\today}

\begin{abstract}
	We consider a Skyrme fluid with a constant radial profile in locally rotational Kantowski-Sachs spacetime.  We choose a suitable change of variable so that the Field equations modified to an autonomous system.  Then we finds the critical points and inspect the stability of them.  Finally, the flow on the Poincar{\'{e}} sphere is shown and consequently the behavior at infinity is determined.  To analyze the non-hyperbolic critical points Center Manifold Theory has been used.  Possible bifurcation scenarios also have been explained.
\end{abstract}

\keywords{ Einstein-Skyrme model; Critical point; Stability; Center Manifold Theory}
\pacs{ 98.80.-k, 95.35.+d, 95.36.+x}

\maketitle

\section{Introduction}	
The Skyrme model is essentially a nonlinear theory of pions, proposed by Skyrme\cite{SkyrmeModel1},\cite{SkyrmeModel2} in the early 60's.  In this model baryons can be treated as topological soliton solutions in the configuration space and it has applications in-various branches of physics.  In general relativity the models with Skyrme field as the matter source are termed as Einstein-Skyrme model.  From the point of view of cosmological scenarios, the Skyrme fluid is an anisotropic fluid without heat flux and the equation of state parameter is constrain as $| \omega_s| \leqslant \frac{1}{3}$.  In 1977, Collins\cite{Collins} first analyzed the model (for the case of perfect fluid) as a two-dimensional dynamical system without cosmological constant.  The author showed that all general relativistic models filled with perfect fluid are geodesically incomplete at past and future.  The fluid energy density becomes infinite at both the singularities.  By transforming the field equations to an autonomous system a dynamical system analysis has been done using Poincar\'{e}-Bendixon Theorem.  L. Parisi et al.\cite{L.Parsi} have studied the dynamical evolution in this model with a cosmological constant with background geometry as locally rotational Kantowski-Sachs(KS) space-time.  In the frame work by dynamical system analysis, they have shown that there are two stable critical points with exponential scale factor for the 2D sphere of the KS model when the radial profile is chosen as constant.  On the other hand for variable radial profile both anisotropic Bianchi I and KS model have been investigated only numerically and for the existence of the solutions there should be restrictions both on the cosmological constant as well as on the Skyrme coupling.  Recently, Paliathanasis et al.\cite{Palia} have considered this Skyrme model in KS space-time for studying quantum cosmology.  Formulating the Whealer-Dewitt equation, they have performed the Lie symmetry classification and have obtained the invariant solutions for the wave function.  The solutions are written in closed form, constructing the Noetherian conservation laws for the field equations.
Moreover, The Einstein-Skyrme model is also used in astrophysical context.  Static black hole solutions having regular event horizon has been formulated for this model.  Such black hole solutions are asymptotically approached to Schwarzschild blackhole and they violates the ``no-hair" conjecture for blackholes.  Recently, Mondal et al.\cite{Mondal} have studied the Einstein-Skyrme model for KS model from the point of view of symmetry analysis.  They have obtained both Lie and Noether symmetry of the cosmological model and are able to obtain cosmological solution by suitable transformation in the augmented space.\\
The present work deals with Einstein-Skyrme model for KS spacetime with constant radial profile and the coupled non-linear modified Einstein Field Equations are analyzed by converting them into an autonomous system.  We first analyze some non-generic critical points in the sense that they are obtained by considering particular values for the parameter on the 2D-hyperplane.  Next we discuss the critical points of 3D-hyperplane using Hartman-Grobman theorem and Center Manifold theory\cite{LPerko}-\cite{Perko}. 
We also discuss the behavior of trajectories `` at infinity '' by using Poincar\'{e} sphere\cite{Poincare}-\cite{Walcher} where we project from the centre of the unit sphere $S^3=\lbrace (X,Y,Z,W)\in \mathbb{R}^3| X^2+Y^2+Z^2+W^2=1\rbrace$ onto the 3D-space tangent to $S^3$ at either the north pole or south pole.  This technique has the advantage that the critical points at infinity proliferate along the equator of the Poincar\'{e} sphere.  Therefore these critical points  are of a simpler to study than the critical points at infinity on 3D-plane\cite{CJF}.   The paper is organized as follows:\\
The basic equations for Einstein-Skyrme model are presented in section 2.  The stability analysis of the equilibrium points of the corresponding autonomous system has been presented in section 3.  The Poincar{\'{e}} sphere and behavior at infinity are also discussed in this section.  
Finally, section 4 gives a discussion of the useful results of this work.

\section{Basic Equations}
In this modified gravity theory, the action takes the form 
\begin{equation} \label{eqn:Action}
I=\int d^4x\sqrt{-g}(R-2\Lambda)+S_l ,
\end{equation}
here $S_l$ is the action for the Skyrme field having expression
\begin{eqnarray} \label{eqn:SkyrmeField}
S_l=\frac{l}{2}\int d^4x\sqrt{-g}(\frac{1}{2}R_\alpha R^\alpha+\frac{\lambda}{16}F_{\alpha\beta}F^{\alpha \beta}) ,
\end{eqnarray}
where $F_{\alpha \beta}=[R_\alpha ,R_\beta]$, $R_\alpha=U^{-1}U$, $U(x^\alpha)$ is the SU(2)-valued scalar field, $\Lambda$ is the usual cosmological constant and $l,\lambda$ are positive parameters.  From action principle, by varying the action with respect to the metric coefficients the modified Einstein field equations for the present model are obtained as,
\begin{equation}  \label{eqn:ModEinsField}
G_{\alpha \beta}+\Lambda g_{\alpha \beta}=lT_{\alpha \beta} ,
\end{equation}
where $T_{\alpha \beta}$ is the energy-momentum tensor for the Skyrme field and $l$ is the usual Einstein coupling constant.  Further, the evolution of the Skyrme field $U(x^\alpha)$ can be obtained by varying the action with respect to it as 
\begin{equation} \label{eqn:EvoSkyrme}
g^{\alpha \beta}(R_{\alpha ; \beta}+\frac{\lambda}{4}([R^{\sigma},F_{\sigma \alpha}])_{; \beta})=0.
\end{equation}
In the present work we shall consider Kontowski-Sachs space-time model having line element
\begin{equation} \label{eqn:Kontowski}
ds^2=-dt^2+A(t)^2dr^2+B(t)^2d\Omega_2 ^2 ,
\end{equation}
where A and B are two scalar factor of the model and $d\Omega_2 ^2=d\theta ^2 + sin^2\theta d\phi ^2$ is the metric on unit 2-sphere.  It should be noted that the present space-time is locally rotational having a four-dimensional killing algebra SO(3) with the killing vector $\partial r$.  Now the scalar field equation (\ref{eqn:EvoSkyrme}) is identically satisfied for constant$(=\frac{\pi}{2}+\sigma \pi ,\sigma \in \mathbb{Z})$ radial profile by choosing 
\begin{eqnarray}  
T_t^t&=T_r^r=-\frac{l}{B^2}(1+\frac{\lambda}{2B^2}), \label{eqn:ChoiceOfRadial1}\\ T_\theta^\theta&=T_\phi^\phi=\frac{l\lambda}{2B^4}.  \label{eqn:ChoiceOfRadial2}
\end{eqnarray}
However, for the present space-time model the explicit form of energy-density, thermodynamic pressure, heat flux and the traceless stress tensor (as measured by the observer $u^{\mu}$) are, 
\begin{eqnarray}  
\rho_s&=\frac{1}{B^2}(\bar l+\frac{\mu}{2B^2}), \label{eqn:ED-TP-HF1}\\
p_s&=-\frac{1}{3B^2}(\bar l-\frac{\mu}{2B^2}) \label{eqn:ED-TP-HF2}\\
and \nonumber \\
q^\mu&=0, \pi_{\mu \gamma}=diag(0,-\frac{2}{3}\rho_s,\frac{1}{3}\rho_s,\frac{1}{3}\rho_s) \label{eqn:ED-TP-HF3},
\end{eqnarray}
where, $\bar{l}=lK$ and $\mu=\bar{l}\lambda$.\\
The equation of state parameter for the Skyrme fluid has the expression,
\begin{equation}
\omega_s=\frac{p_s}{\rho_s}=-\frac{1}{3}(\frac{\bar{l}B^2-\mu}{\bar{l}B^2+\mu})
\end{equation}
which gives, $\omega_s\approx-\frac{1}{3}$ when $\bar{l}B^2>>\mu$ and $\omega_s\approx \frac{1}{3}$ when $\bar{l}B^2<<\mu$ i.e. $-\frac{1}{3}\leqslant \omega_s \leqslant \frac{1}{3}$.\\
Further, for line-element (\ref{eqn:Kontowski}), the modified field equation (\ref{eqn:ModEinsField}) is expressed in terms of the parameters, $\lbrace A, B \rbrace$ (reference \cite{Palia}),

\begin{eqnarray}
2\frac{\dot{A}\dot{B}}{AB}+\frac{\dot{B}^2}{B}+\frac{1}{B^2}-\Lambda&=\frac{1}{B^2}(\bar{l}+\frac{\mu}{2B^2})\\
2\frac{\ddot{B}}{B}+\frac{\dot{B}^2}{B^2}+\frac{1}{B^2}-\Lambda&=\frac{1}{B^2}(\bar{l}+\frac{\mu}{2B^2})\\
\frac{\ddot{A}}{A}+\frac{\ddot{B}}{B}+\frac{\dot{A}\dot{B}}{AB}-\Lambda&=-\frac{\mu}{2B^4}
\end{eqnarray}
Now by suitable choice of variable
\begin{equation}
\frac{\dot{A}}{A}=x,\frac{\dot{B}}{B}=y,B=z.\\
\end{equation}
We get the autonomous system
\begin{eqnarray}
\dot{x}&=\Lambda-x^2-2xy-\frac{\mu}{2z^4} \label{eqn:Auto3D1}\\
\dot{y}&=xy-y^2  \label{eqn:Auto3D2}\\
\dot{z}&=yz    \label{eqn:Auto3D3}
\end{eqnarray}
The above autonomous system is $C^1(\mathbb{R}^3-XYplane)$ and `dot' represents the differentiation with respect to t.\\
Owing to change of variable the equation of state parameter takes the form
\begin{equation}
w_s= -\frac{1}{3}(\frac{\bar{l}z^2-\mu}{\bar{l}z^2+\mu}).
\end{equation}
The Hubble parameter (H) can be expressed as
\begin{eqnarray}
H&=\frac{1}{3}(\frac{\dot{A}}{A}+2\frac{\dot{B}}{B}) \nonumber\\
&=\frac{1}{3}(x+2y).
\end{eqnarray}
The density parameter for the scalar field has the expression
\begin{eqnarray}
\Omega_s&=\frac{\rho_s}{3H^2}  \nonumber\\
&=\frac{3}{2}(\frac{2\bar{l}z^2+\mu}{z^4(x+2y)^2})
\end{eqnarray}
The deceleration parameter (q) is a dimensionless measure of the cosmic acceleration defined as
\begin{eqnarray}
q&=-(1+\frac{\dot{H}}{H^2}) \nonumber \\
&=-(1+\frac{3(\Lambda-x^2-2y^2-\frac{\mu}{2z^2})}{(x+2y)^2}).
\end{eqnarray}

In the following section we shall determine the critical (equilibrium) points and analyze the stability criteria for the critical points (CPs). 

\section{Stability Analysis} \label{StabilityAnalysis}
For global stability analysis we need to analyze the behavior of the critical points on the hyperplanes parallel to the XY, YZ and XZ plane respectively.  Initially we intensify the global behavior of the vector field on the hyperplanes.  The flow of the 3D autonomous system (\ref{eqn:Auto3D1}-\ref{eqn:Auto3D3}) carries the behavior of the vector field of 2D autonomous systems.  
\subsection{The vector fields on the 2D autonomous system}
Let us first consider B(t) is constant.  For  z is a non zero constant (c) the system (\ref{eqn:Auto3D1}-\ref{eqn:Auto3D3}) turns into an autonomous system as follows,
	 \begin{equation}
		\begin{aligned}
		\dot{x}&=k-x^2\\
		\dot{y}&=0\\
		\end{aligned}\label{equn2Da}
		\end{equation}
		where $k=\Lambda-\frac{\mu}{2c^4}$, a constant.

\begin{figure}[h!]
			\centering
			\includegraphics[scale=.4]{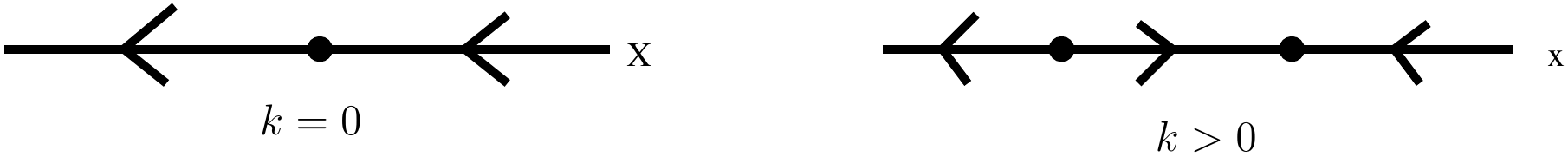}
			\caption{\label{fig:z_constant}Stability for k=0 and $k>0$.}
			\label{fig:x1}
	\end{figure}

For $k>0$ there are two critical points at $x=\pm \sqrt{k}$ and corresponding eigenvalues of the Jacobian matrix are $\mp 2 \sqrt{k}$.  So $x=\sqrt{k}$ is stable while the critical point at $x=-\sqrt{k}$ is unstable.  For k=0, the flow determined by $\dot{x}=-x^2$ (\ref{fig:z_constant}). For $k<0$ there are no critical points.

\begin{figure}[h!]
	\centering
	\includegraphics[scale=.4]{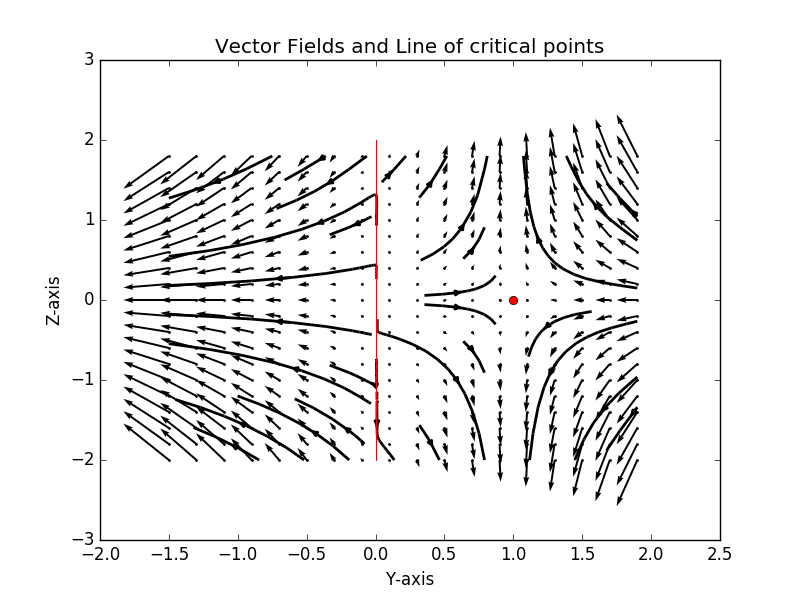}
	\caption{The local phase portrait for p=1.  Red vertical-line shows the line of critical points.}
	\label{fig:2d}
\end{figure}

Next we analyze the flow on the hyperplane which is parallel to YZ-plane i.e. x is constant= p(say). On this hyperplane the system (\ref{eqn:Auto3D1}-\ref{eqn:Auto3D3}) turns into the 2D autonomous system as follows.
\begin{eqnarray}
\dot{y}&=y(p-y) \label{eqn:Auto2Db1}\\
\dot{z}&=yz    \label{eqn:Auto2Db2}
\end{eqnarray}
This autonomous system has a critical point (p,0) and a line of critical point $(0,z_c)$ where $z_c\in \mathbb{R}-\lbrace0\rbrace$.  The flow near (p,0) behaves like a saddle.  The critical points $(0,z_c)$ are non-hyperbolic critical points.  As the center manifold near the points  $(0,z_c)$ is y=0, so we can not get any flow along Z-axis.  With the aim to draw the phase portrait using Euler method for numerical solution of differential equation,  we get the figure \ref{fig:2d} which shows the line of critical points $(0,x_c)$ are unstable for p=1.  The values of the cosmological parameters at the critical points are shown in the following table (\ref{table2}).

\begin{table} 
	\caption{Critical points and corresponding values of cosmological parameters.}
	\begin{ruledtabular}
		\begin{tabular}{ c|ccc }
			\hline
			CPs  & H &  q  & Eig.Vals. \\
			\hline
			$y=p,z=0$&p&$-1$&$\lbrace$-p,p$\rbrace$\\
			\hline
			$y=0,z=z_c$ &$\frac{1}{3}$p&$-1$&$\lbrace$p,0$\rbrace$
		\end{tabular}\label{table2}
	\end{ruledtabular}
\end{table}	
Finally we find that there is no critical point on the plane parallel (not equal) to XZ-plane.  Only XZ-plane contains critical points that are exactly identical to the critical points of 3D autonomous system (\ref{eqn:Auto3D1}-\ref{eqn:Auto3D3}).  So the behavior of the vector field in a small neighborhood of these critical points has been explicated following.

\subsection{Vector fields in the 3D autonomous system}
In 3D autonomous system (\ref{eqn:Auto3D1}-\ref{eqn:Auto3D3}) the  critical points (CPs)  are $(x_c,0,z_c) \in \mathbb{R}^3$ where $x_c=\pm\frac{\sqrt{\Lambda z_c^4-\mu}}{z_c^2}$ and $z_c\neq0$.\\
The Jacobian Matrix is
\[
J(x,y,z)=
\left[{\begin{array}{ccc}
	
	-2x-2y & -2x & \frac{2\mu}{z^2}\\
	y & x-2y & 0\\
	0 & z & y
	\end{array}} \right]
\]
at any critical points $(x_c,0,z_c)$, 
\[
J(x_c,0,z_c)=
\left[{\begin{array}{ccc}
	-2x_c & -2x_c & \frac{2\mu}{z_c^2}\\
	0 & x_c & 0\\
	0 & z_c & 0
	\end{array}}\right]
\]
The eigenvalues of Jacobian matrix and the values of cosmological parameters at critical point $(x_c,0,z_c)$ are shown in Table (\ref{table3}).  In this table, the eigenvalues of $J(x_c,0,z_c)$ are  $0,x_c$,$-2x_c$.  One should note that, all the critical points of the system (\ref{eqn:Auto3D1}-\ref{eqn:Auto3D3}) are non-hyperbolic (NH CPs) in nature.  The zero eigenvalue of Jacobian matrix makes Hartman-Grobman theorem unbefitting.  Also the Liapunov function is difficult to find for the nonlinear system (\ref{eqn:Auto3D1}-\ref{eqn:Auto3D3}).  Center Manifold theory (CMT) extricates from such a situation.
\begin{table} 
	\caption{Critical points and corresponding values of cosmological parameters.}
	\begin{ruledtabular}
		\begin{tabular}{ c|ccccc }
			\hline
			CPs & H &q&$\Omega_s$&$\omega_s$& Eig.Vals.\\
			\hline
			$x=x_c,y=0,z=z_c$,\\ $x_c=\pm\frac{\sqrt{\Lambda z_c^4-\mu}}{z_c^2}$ and $z_c\neq 0. $&$-\frac{1}{3}x_c$&$-1$&$ \frac{2\bar{l}z_c^2+\mu}{2\Lambda z_c^4 - \mu}$&$ -\frac{1}{3}\frac{\bar{l}z_c^2-\mu}{\bar{l}z_c^2+\mu}$&$\lbrace 0,x_c,-2x_c \rbrace$\\
		\end{tabular}\label{table3}
	\end{ruledtabular}
\end{table}	

It is convenient to find the center manifold at the origin.  So we shift the point $(x_c,0,z_c)$ to the origin and find the center manifold at origin.  The system of equations (\ref{eqn:Auto3D1}-\ref{eqn:Auto3D3}) in the shifted coordinate system takes the form as follows.
\begin{eqnarray}
\dot{x}&=\Lambda-(x+x_c)^2-2(x+x_c)y-\frac{\mu}{2(z+z_c)^4} \label{eqn:AutoShift3D1}\\
\dot{y}&=(x+x_c)y-y^2    \label{eqn:AutoShift3D2}\\
\dot{z}&=y(z+z_c)       \label{eqn:AutoShift3D3}
\end{eqnarray}

The Jordan form of $J_{Jordan}(x_c,0,z_c)$ (or $J_{Jordan}(0,0,0)$ in the shifted co-ordinate system) is

\[
J(x_c,0,z_c)=
\left[{\begin{array}{ccc}
	-2x_c & 0 & 0 \\
	0 & x_c & 0\\
	0 & 0 & 0
	\end{array}}\right]
\]
where the change of basis matrix is 
\[
P=
\left[{\begin{array}{ccc}
	1 & -2(\frac{1}{3}-\frac{\mu}{3x_c^2z_c^4}) & \frac{\mu}{x_cz_c^5} \\
	0 & 1 & 0\\
	0 & \frac{z_c}{x_c} & 1
	\end{array}}\right]
\]
where $det~P=1$.\\
After change of standard basis in $\mathbb{R}^3$ by P, the system (\ref{eqn:AutoShift3D1}-\ref{eqn:AutoShift3D3}) changes as follows.
\begin{widetext}
\begin{eqnarray}
\dot{X}&=-2x_cX-X^2+(-\frac{2}{3}-\frac{7\mu}{x_c^2z_c^4}+\frac{2\mu^2}{3x_c^4z_c^8}+\frac{z_c^2}{x_c^2})Y^2+(\frac{\mu^2}{x_c^2z_c^{10}}-\frac{5\mu}{z_c^6})Z^2-(\frac{\mu}{x_c^2z_c^4})XY +\frac{\mu}{x_cz_c^5}(-11 -\frac{2\mu}{x_cz_c^4}+\frac{\mu}{x_c^2z_c^4})YZ-(\frac{2\mu}{x_cz_c^5})ZX \label{eqn:AutoShift3D_J1}\\
\dot{Y}&=x_cY-Y^2+XY+(-\frac{2}{3}+\frac{2\mu}{3x_c^2z_c^4})Y^2+(\frac{\mu}{x_cz_c^5})ZY \label{eqn:AutoShift3D_J2}\\
\dot{Z}&=-\frac{z_c}{x_c}XY+(-\frac{5}{3}+\frac{z_c}{x_c}+\frac{2\mu}{3x_c^2z_c^4})Y^2+(1+\frac{\mu}{x_cz_c^5})ZY \label{eqn:AutoShift3D_J3}
\end{eqnarray}
\end{widetext}

Center manifold of the above system (\ref{eqn:AutoShift3D_J1}-\ref{eqn:AutoShift3D_J3}) is 
\begin{eqnarray}
X&=(\frac{\mu^2}{2x_c^3z_c^{10}}-\frac{5\mu}{2x_cz_c^6})Z^2 + \textrm{higher order terms}\\
Y&=0
\end{eqnarray}
The flow along the center manifold (we get from Eqn. (\ref{eqn:AutoShift3D_J3})) is $\dot{Z}=0$.
\begin{figure}[t]
	\centering
	\includegraphics[scale=.3]{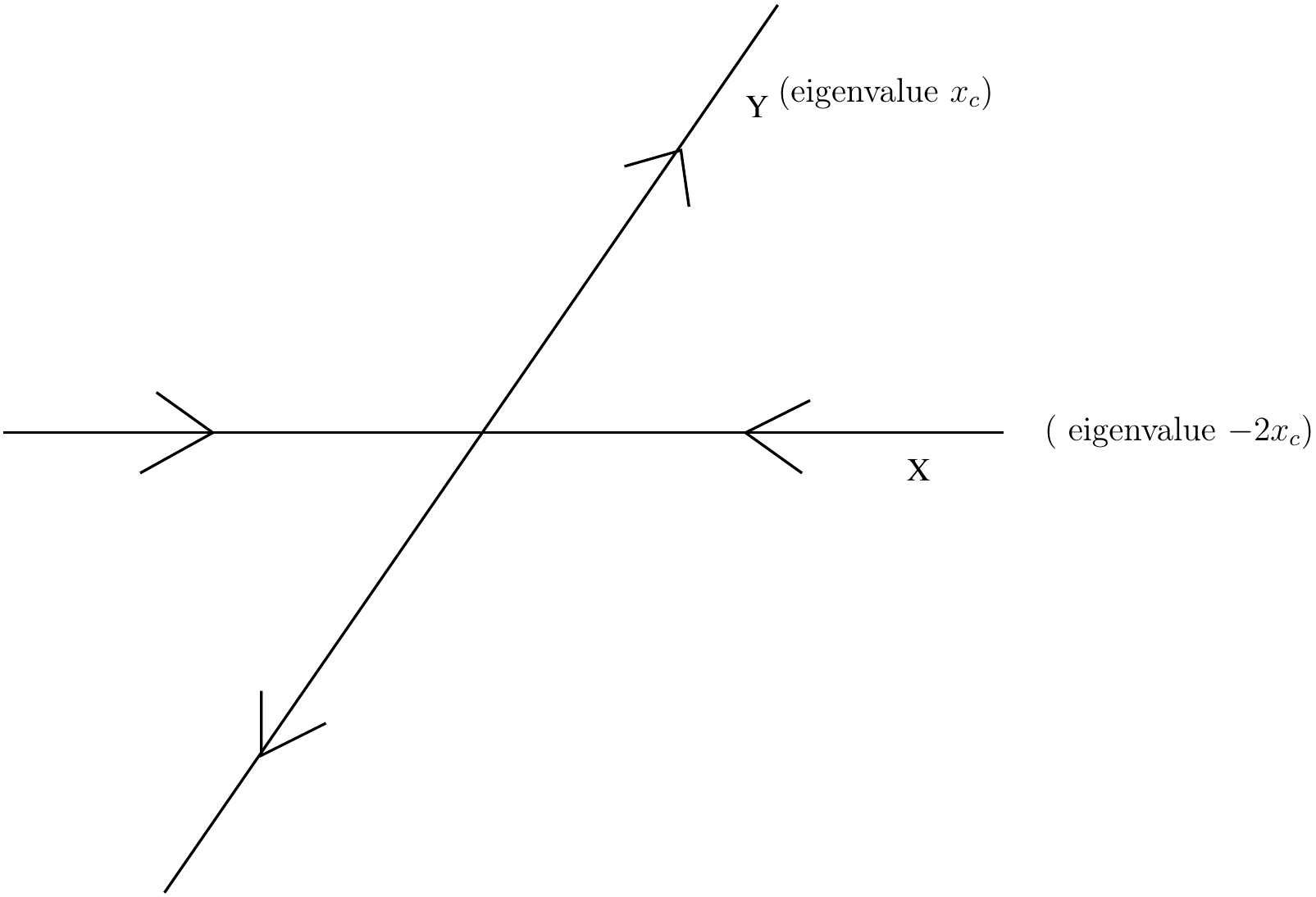}
	\caption{The local phase portrait for the system when $x_c>0$.}
	\label{fig3D:2a}
\end{figure}
\begin{figure}[t]
	\centering
	\includegraphics[scale=.3]{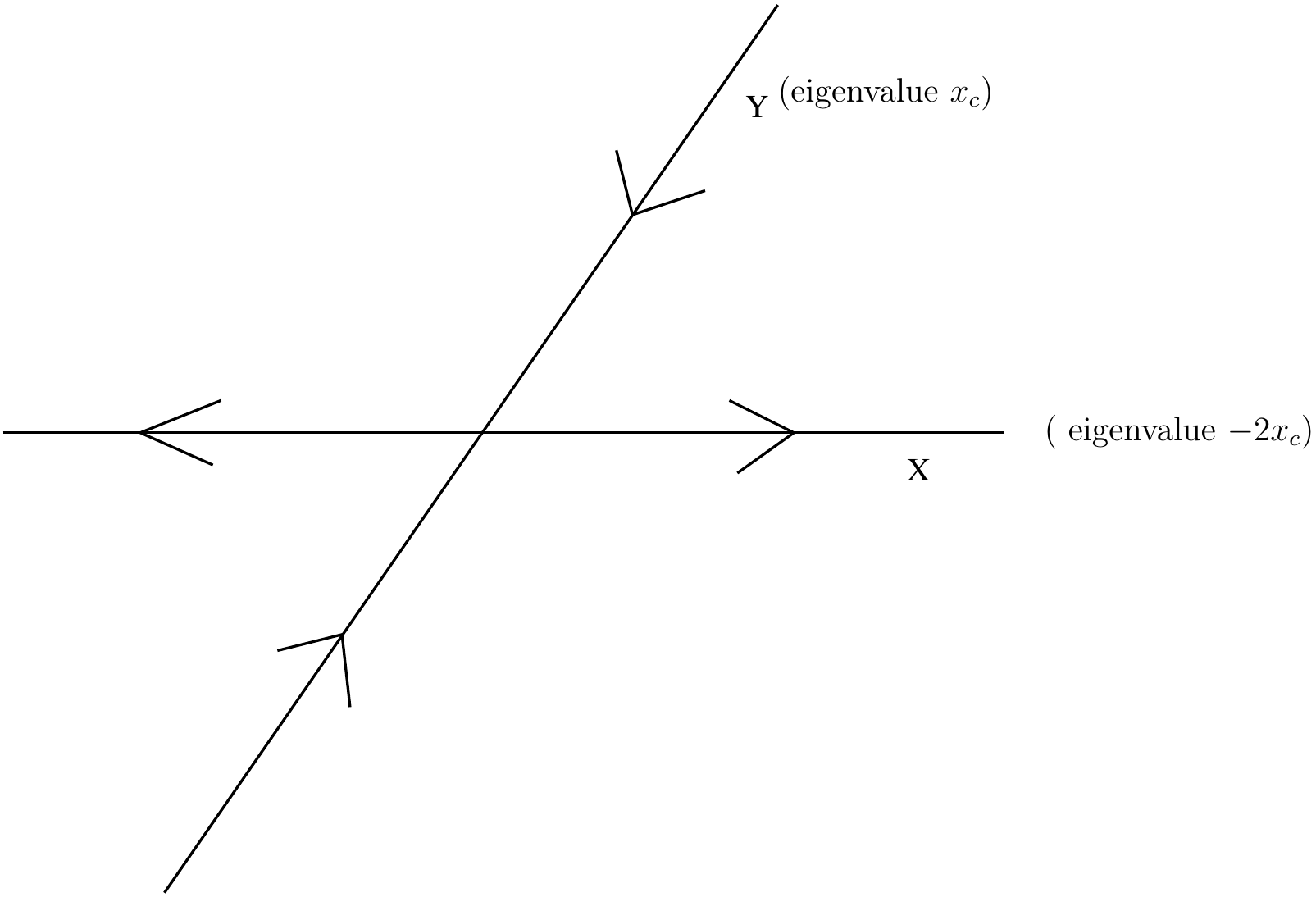}
	\caption{The local phase portrait for the system when $x_c<0$.}
	\label{fig3D:2b}
\end{figure}
The flow is parallel to XY plane as shown in the figure (\ref{fig3D:2a}) (when $x_c>0$).  In a small neighborhood of the point $(x_c,0,z_c)$ the flow is unstable and asymptotic to Y-axis as t goes to infinity.  If we confine the neighborhood parallel to the XY plane, the flow is unstable near the CP and behaves like saddle.  On the other hand, when $x_c<0$, the flow near the CP is shown in the figure (\ref{fig3D:2b}).  The flow is unstable near CP and asymptotic to X-axis as time moves on. 

\subsection{Poincar\'{e} sphere and the Behavior at infinity}
The local behavior of 2D and 3D autonomous system do not give any idea about the behavior at infinity of the system (\ref{eqn:Auto3D1}-\ref{eqn:Auto3D3}).  
To find the global behavior we initially see the autonomous system (\ref{eqn:Auto3D1}-\ref{eqn:Auto3D3}) onto the sphere $S^3$ which is embedded in $\mathbb{R}^4$.  Next we project the upper hemisphere (suitably chosen) of $S^3$ onto $\mathbb{R}^3$.  For convenience (to get $C^1(\mathbb{R}^3)$ system) we take $\mu=0$ in system (\ref{eqn:Auto3D1}-\ref{eqn:Auto3D3}).\\
We use the transformation of coordinate chosen by (reference \cite{Kappos}, \cite{Walcher}, \cite{Roland} and section 3.10 in \cite{LPerko})
\begin{equation}
x=\frac{X}{W},\ y=\frac{Y}{W},\ z=\frac{Z}{W}
\end{equation}
for $V=(X,Y,Z,W)\in S^3$ where $|V|=1$ and $v=(x,y,z)\in \mathbb{R}^3$.  Now the critical points at infinity lie on the equator of the Poincar\'{e} sphere $S^3$ at points $(X, Y, Z, 0)$ with $X^2+Y^2+Z^2=1$ and satisfy the following equations
\begin{eqnarray}
XY(2X+Y)=0\\
XZ(3Y+X)=0\\
ZY(2Y-X)=0
\end{eqnarray} 
So the critical points at infinity are at $A(1,0,0,0)$, $A'(-1,0,0,0)$, $B(0,1,0,0)$, $B'(0,-1,0,0)$, $C(0,0, 1,0)$, $C'(0,0,-1,0)$.  The flow in the neighborhood of $(\pm 1,0,0,0)$ is determined by
\begin{eqnarray}
\pm \dot{y}&=-2y-y^2+\Lambda yw^2\\
\pm \dot{z}&=-z-3yz+\Lambda zw^2\\
\pm \dot{w}&=-w-2yw+\Lambda w^3
\label{equation3D3}
\end{eqnarray}
The eigenvalues of Jacobian matrix (at CPs) are $\lbrace \pm 2,\pm 1,\pm 1 \rbrace$.  So the critical points $(\pm 1,0,0,0)$ is unstable or stable node.  The flow near the point $A$ is topologically equivalent to the flow near $A'$ with the reverse direction as the highest degree of the system (\ref{eqn:Auto3D1}-\ref{eqn:Auto3D3}) is even.  However, for $x_c>0$ the flow is stable along $[1~0~0]^T$.  So, $A$ is unstable node and $A'$ is stable node.\\
The flow in the neighborhood of $(0,1,0,0)$ is determined by
\begin{eqnarray}
\pm \dot{x}&=x+2x^2+\Lambda w^2\\
\pm \dot{z}&=-2z+xz\\
\pm \dot{w}&=-w+xw
\end{eqnarray}  
The eigenvalues of Jacobian matrix (at CP) are $\lbrace -1,2,1 \rbrace$ or $\lbrace 1,-2,-1 \rbrace$.  So the critical points $B$ and $B'$ are saddle.\\
Similarly, the flow in the neighborhood of $(0,0,\pm 1,0)$ is determined by
\begin{eqnarray}
\pm \dot{x}&=3xy+x^2-\Lambda w^2\\
\pm \dot{y}&=2y^2-xy\\
\pm \dot{w}&=yw
\end{eqnarray}
As the eigenvalues of Jacobian matrix (at CP) are all zero so Center Manifold Theory is not applicable.  The flow along the X-axis at (0,0,1,0) is given by $\dot{x}=-x^2$ and along Y-axis given by $\dot{y}=-2y^2$ and at (0,0,-1,0) the flow behaves same but orientation becomes opposite.  The behavior at infinity given by the behavior of flow on the equator of $S^3$ that is $S^2$.

\section{Bifurcation Analysis}

In system (\ref{equn2Da}), the vector field $f(x)=-x^2$ is structurally unstable and k=0 is a bifurcation value.

\begin{figure}[h!]
			\centering
			\includegraphics[scale=.4]{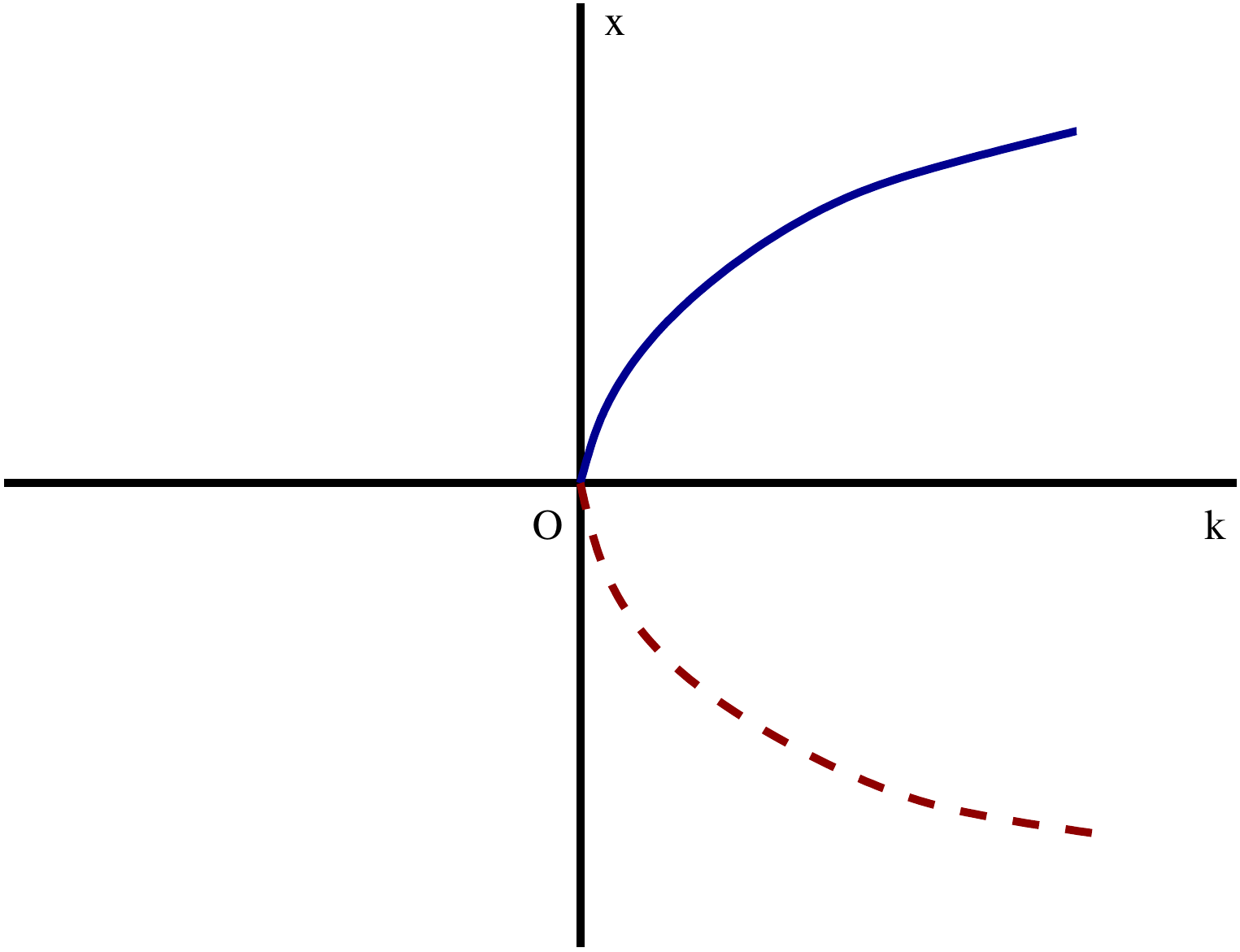}
			\caption{The saddle-node bifurcation diagram.}
			\label{fig:x1}
	\end{figure}

For 3D autonomous system $\Lambda=\mu=0$ is a bifurcation value.  Also phase-transition of the universe occur  at this bifurcation value.  

\begin{table} 
	\caption{Critical points and corresponding values of cosmological parameters.}
		\begin{tabular}{ ||c||c||c||c|| }
			\hline
			CPs & H &q& Eig.Vals.\\
			\hline
			$x=x_c,y=0,z=z_c$,\\ $x_c=\pm\frac{\sqrt{\Lambda z_c^4-\mu}}{z_c^2}$ and $z_c\neq 0. $&$-\frac{1}{3}x_c$&$-1$&$\lbrace 0,x_c,-2x_c \rbrace$\\
			\hline
		\end{tabular}\label{table3}
\end{table}

\section{Discussion}
In this work we have studied Einstein-Skyrme model in a Kantowski-Sachs spacetime.  At first the system of equations are transformed to an autonomous system by suitably choosing change of variable and then globally characterize the critical points.\\

An unstable fixed point may be considered as an initial position of the universe and a small perturbation scamper the start of the trajectory of the universe.  The initial position of the universe drives a unique trajectory which starts from an arbitrary close neighborhood  of an unstable critical point and stops to an arbitrary close neighborhood of a stable critical point or goes to infinity.  The qualitative global behavior of the trajectory of the universe is described by inspecting the local behavior of critical points.\\

For convenience we first study the dynamical behavior of the system (\ref{eqn:Auto3D1}-\ref{eqn:Auto3D3}) on the hyperplanes parallel to the XY-plane (when z=constant$\neq 0$), YZ-plane (when x=constant) and ZX-plane (when y=constant) respectively.   We named the planes as A type, B type, C type respectively.\\
It is interesting to note that, A type contains four critical points for $k>0$ and one critical point for k=0.  In the Table 1. the top two points are saddle and when the universe is very close neighborhood of these points it goes away from these critical points but only a trajectory may connect those points as shown in the figure 2 (sky-colored and blue-colored points are connected by a trajectory).  This situation happen when the two saddle points share a trajectory connect the eigenvector of positive eigenvalue of a point with eigenvector of negative eigenvalue of another point.  Third critical point of Table 1. is a stable node which is the final destination of some trajectories.  The fourth critical point is unstable node and the trajectory starting very close to this point either goes towards the third critical point or goes to infinity as time moves on.  Further, the first and third critical points of table 1 represent an accelerated expansion of the universe similar to the cosmological constant (deSitter).  On the other hand, the other two critical points of table 1 represent contracting model of the Universe where the rate of contraction is accelerating.  It is to be noted that for k=0 all the critical points coincide with no expansion or contraction, i.e. static model of the Universe which is not of much interest in the present context.\\
On the B type hyperplanes there is a line of critical points and a saddle point (p,0).  There is no flow along Y-axis on this plane but there is a obvious movement on 3D space along Y-axis.  When the universe is very close to (p,0) at some time, then it goes away from the point to infinity asymptotically to Y-axis.  The critical points in table 2 represent expanding and contracting model of the Universe depending on positive and negative values of p respectively.\\
All the critical points of C type hyperplanes lie on the ZX-plane and the critical points are exactly same as of 3D system of equations.  In the 3D system of equation, the critical points of positive x-coordinate is stable on the XZ-plane and unstable along Y-axis.  Thus for any initial point of the universe in a very small neighborhood of the critical point(x-coordinate positive) goes away from the critical point asymptotically to Y-axis.  Similarly, any initial point of the universe near to a critical point (x-coordinate negative) goes away from the critical point asymptotically to ZX-plane.  The critical point in table 3 represents an expanding or contraction model of the Universe depending on the sign of $x_c$.  This model of the Universe is also deSitter type.  Here the equation of the state parameter for the symmetric fluid is in the phantom region if $lz_c^2 \geqslant \mu$.  Otherwise the Skyrme fluid will be the normal fluid.


\begin{acknowledgements}
The author S. Mishra is grateful to CSIR, Govt. of India for giving Junior Research Fellowship (CSIR Award No: 09/096 (0890)/2017- EMR - I) for the Ph.D work. \\
Conflict of Interest: The authors declare that they have no conflict of interest.
\end{acknowledgements}

\end{document}